\documentclass[12pt]{article}
\usepackage{graphicx}
\usepackage[cp1251]{inputenc}
\begin{document}
\pagestyle{empty} \pagestyle{headings} {\title{Prediction of solar
flares using neutrino detectors of the second generation}}
\author{O.M.Boyarkin$^a$\thanks{E-mail:oboyarkin@tut.by},
\ I.O.Boyarkina\thanks{E-mail:estel20@mail.ru}\ $^b$\\
$^a$\small{\it{Belarusian State University,}}\\
\small{\it{Dolgobrodskaya Street 23, Minsk, 220070, Belarus}}\\
$^b$\small{\it{The University of Tuscia}}\\
\small{\it{DEIM, 47, Paradise str, Viterbo, Italy}}}
\date{}
\maketitle

\begin{abstract}
The physics-based method of forecasting the high energy solar
flares (SFs) with the help of the neutrino detectors utilizing the
coherent elastic neutrino-nucleus scattering (CE$\nu$NS) is
considered. The behavior of the neutrino beams traveling through
the coupled sunspots (CSs) being the sources of the future SFs is
investigated. Two neutrino beams are included into consideration,
namely, the $\nu_{eL}$ beam and the $\nu_{\mu L}$ beam which have
been produced after the passage of the Micheev-Smirnov-Wolfenstein
resonance. It is assumed that the neutrinos possess the charge
radius, the magnetic and anapole moments while the CS magnetic
field is vortex, nonhomogeneous and has twisting. Estimations of
the weakening of the neutrino beams after traversing the resonant
layers are given. It is demonstrated, that this weakening could be
registered by the detectors employing CE$\nu$NS only when the
neutrinos have the Dirac nature. \\[1mm]
PACS number(s): 12.60.Cn, 14.60.Pg, 96.60.Kx, 95.85.Qx, 96.60.Rd.
\end{abstract}
\hspace{1mm}Keys words: Solar flares, forecasting the flares,
neutrino oscillations, magnetic moment, anapole moment, charge
neutrino radius, neutrino detectors, coherent elastic
neutrino-nucleus scattering, RED-100.

\section{Introduction}

Solar flares (SFs) are the most striking explosive form of solar
activity. They take place in the solar atmosphere and release a
wealth of energy, which could be as large as $10^{28}-10^{33}$
erg. Moreover, the super SFs with the energy of $10^{36}$ erg and
more are also possible in the solar conditions \cite{MLA}. The SFs
are often (but not always) followed by coronal mass ejections
which represent eruptions of the solar coronal plasma into the
interplanetary space. It is clear that the high energy SFs are
very destructive when they are focussed on the Earth as it was in
1859 (Carrington event \cite{RAC}). It might be worth pointing out
that the flares may be also occurred in other Sun-like stars.
Flares at these stars are also dangerous for crew members of
interplanetary spaceships. Therefore, for our ever more
technologically dependent society forecasting the SFs has a large
practical value.

It is generally accepted that the magnetic field is the basic
energy source of the SF \cite{SI81,KST11}. The course of the
periods of the high solar activity, the magnetic flux
$\sim10^{24}\ \mbox{G}\cdot\mbox{cm}^2$ \cite{DJ81} is erupted
from the solar center and stored  on the sunspots. In so doing,
big sunspots of opposite polarity could be paired forming, the
so-called, coupled sunspots (CSs). Then accumulation of the
magnetic energy  starts. The more energetic the SF, the more will
be the magnetic field strength of the CS. For example, in the case
of the super SFs $B_{cs}$ may reach the values of $10^8$ G and
upwards. The length of the initial SF stage is extended from
several to dozens of hours. Obviously, the successful SF
forecasting should be based on the analysis of the phenomena
occurring in the CS region at the initial stage of the SF.
Previous studies of forecasting the SF are carried out with a help
of $\gamma$-telescopes which observe the Sun collecting particle
measurements related to SFs. Further, this huge amount of
observations is transferred, stored, and handled. To deal with
this data, a new method of Machine Learning (ML) has been created.
The ML method has been utilizing such as models, as support vector
machines \cite{RQA}, neural networks \cite{OWA}, a regression
model \cite{JYL}, an extremely randomized trees \cite{NNI} and so
on. It was proposed to use satellites in the solar orbit with
built in the ML capability that continuously monitor the Sun.
These observatory will use the ML to calculate the probability of
the solar explosions from the remote sensing data. Currently in
operation are the following space-borne instruments: Solar and
Heliospheric Observatory, Solar Dynamics Observatory, Advanced
Composition Explorer, Atmospheric Imaging Assembly, Large Angle
Spectroscopic Coronagraph on the Solar and Heliospheric
Observatory. The ML can clarify which feature is most effective
for forecasting the SFs. However, to date, it is not known which
of the models used in the ML is the best.

Among the physics-based models that are used for forecasting the
SF, the so-called kappa scheme proposed by a team of Japanese
physicists \cite{KK20} should be noted. Their model forecasts the
high energy SFs through a critical condition of
magnetohydrodynamic instability, triggered by magnetic
reconnection. The group tested the method using observations of
the Sun from 2008 to 2019. In the most cases, the method correctly
identifies which regions will produce the high energy SF within
the next 20 hours, The method also provides the exact location
where each the SF will begin and limits on how powerful it will
be.

The Sun is not only the source of electromagnetic radiation, it
also emits a huge stream of electron neutrinos
($N_{\nu_{eL}}\simeq6\times10^{10}\mbox{cm}^{-2}\mbox{s}^{-1}$).
It is obvious that with the help of sensitive neutrino detectors
it will be possible to obtain information about events occurring
in the CS region. It could be done with the detectors of the
second generation whose work is based on coherent elastic
(anti)neutrino-atomic nucleus scattering (CE$\nu$NS). This type of
low-energy (anti)neutrino interaction was predicted in 1974
\cite{KF74,FD74} and was recently discovered by COHERENT
Collaboration \cite{AK17}. It was shown that neutrinos and
antineutrinos of all types can interact by exchanging the
$Z$-boson with the atomic nucleus as a whole, i.e. coherently.
This take place with the (anti)neutrino energy being less than 50
MeV when De Broglie wavelength increases to the value of order of
the nucleus charge radius $R=1.12\times(A)^{1/3}10^{-13}$ cm ($A$
is the number of nucleons). The cross section of the CE$\nu$NS is
described by the formula
$$\sigma\simeq\mbox{few}\times10^{-45}N^2(E_{\nu})^2\ \mbox{cm}^2,$$
where $N$ is the number of neutrons, $E_{\nu}$ is the
(anti)neutrino energy, expressed in MeV. Thanks to the $N^2$
factor, the cross section of this process is large, it is more
than two orders of magnitude (for heavy nuclei) larger than the
cross section of other known processes describing the interactions
of low-energy (anti)neutrinos. To satisfy the demands for a
coherently enhanced interaction, (anti)neutrinos need to be have
energies in the MeV-regime. As of now, the two favored
antineutrino sources are nuclear reactors (Connie, Conus,
Ncc-1701) and $\pi$DAR sources (Coherent, Ccm, Ess). These
antineutrino sources taking together allow us to investigate
different aspects of CE$\nu$NS at various energies and neutrino
flavors.

Detectors based on the employment of the CE$\nu$NS are already
being used for monitoring the operation of a nuclear reactor in
the on-line regime. Examples are found in Russian Emission
Detector-100 (RED-100) at Kalininskaya nuclear power plant
\cite{AKIM}. Installed at a distance of 19 meters from a nuclear
reactor, where the reactor antineutrino flux reaches the values
$1.35\times10^{13}\ \mbox{cm}^{-2}\ \mbox{c}^{-1}$, RED-100 should
record 3300 antineutrino events per day. Moreover, in the future,
it is planned to scale the detector by a factor of 10 to the mass
of the sensitive volume of the order of 1 ton (RED-1000)
\cite{AYA1}. This will make it possible to register 33,000 events
per day. Obviously, such detectors can be also developed for
observations of solar neutrinos (but here we shall deal with the
coherent elastic neutrino-atomic nucleus scattering). Then, if to
study the behavior of the solar $pp$ neutrinos the detector
similar in design to the RED-1000 will be used, then it could
register about 2000 neutrino events per day.

The aim of our work is to investigate the possibility of
forecasting the high energy SF with the help of the neutrino
detectors utilizing the CE$\nu$NS. The work represents
continuation of the papers \cite{BB200,BGAL,BGAL1,Neut} in which
the correlation between the SF and behavior of the electron
neutrino beam in the CS magnetic field during the initial stage of
the SF was discussed. In contrast to the previous works, we take
into account all the neutrino multipole moments and carry out the
analysis for the Dirac and Majorana neutrinos. In the next Section
constraining by two flavor approximation we obtain the evolution
equation and find all the resonance conversions both of the
electron neutrinos and of the muon neutrinos that emerged from the
MSW resonance in the convective zone of the Sun. Further we give
estimations of weakening the neutrino beam after traversing the
resonant layers and demonstrate that this quantity could be
observed by the neutrino detector of the second generation.
Finally, in Section 3, some conclusions are drawn.

\section{Neutrino behavior in solar matter}

In the standard model (SM) the neutrino magnetic moment is
determined by the expression
$$\mu_{\nu}=10^{-19}\mu_B
\Big(\frac{m_{\nu}}{\mbox{eV}}\Big).\eqno(1)$$ It clear that such
a small value cannot bring to any observable effects in real
magnetic fields. Hence, if we use the values of the neutrino MMs
being close to the upper experimental limits $(10^{-10} -
10^{-11})\mu_B$, then we should go beyond the SM. As an example of
such a SM extension we may employ the left-right symmetric model
which is based on the $SU(2)_R\times SU(2)_L\times U(1)_{B-L}$
gauge group \cite{ICP74, RNM75,GSRN}.

In the one-photon approximation, the effective interaction
Hamiltonian satisfying the demands both of the  Lorentz and of the
electromagnetic gauge invariance is determined by the following
expression \cite{Ni82,Ka82}
$${\cal{H}}^{(\nu)}_{em}(x)=
\sum_{i,f}\overline{\nu}_i(x)
\{i\sigma_{\mu\lambda}q^{\lambda}[F^{if}_M(q^2)+iF^{if}_E(q^2)\gamma_5]+
(\gamma_{\mu}-q_{\mu}q^{\lambda}
\gamma_{\lambda}/q^2)[F^{if}_Q(q^2)+$$
$$+F^{if}_A(q^2)q^2\gamma_5]\}\nu_f(x)A^{\mu}(x),\eqno(2)$$
where $q_{\mu}=p_{\mu}^{\prime}-p_{\mu}$ is the transferred
4-momentum, while $F^{if}_Q, F^{if}_M, F^{if}_E,$ and $F^{if}_A$
are the charge, dipole magnetic, dipole electric, and anapole
neutrino form factors. The form-factors with $i=f$ ($i\neq f$) are
named "diagonal" ("off-diagonal" or "transition") ones. In the
static limit ($q^2=0$), $F^{if}_M(q^2)$, $F^{if}_E(q^2)$ and
$F^{if}_A(q^2)$ determine the dipole magnetic, dipole electric and
anapole moments, respectively. Note, the second term in the
expansion of the $F_Q^{if}(q^2)$ in series of powers of $q^2$
determines the neutrino charge radius
$$<r^2_{if}>=6\frac{dF^{if}_Q(q^2)}{dq^2}\Bigg|_{q^2=0}.\eqno(3)$$
In what follows, we shall be interested in the magnetic moments
(MM), the anapole moments (AM) and the charge radii (NCR).

The exhibiting of neutrino MMs are being searched in the reactors
(MUNU, TEXONO and GEMMA), the accelerators (LSND), and solar
(Super-Kamiokande and Borexino) experiments. The current best
sensitivity limits on the diagonal MMs gotten in laboratory
experiments are as follows
$$\mu_{ee}^{exp}\leq2.9\times10^{-11}\mu_B, \qquad90\%
\ C.L.\qquad [\mbox{GEMMA}] \ \cite{AGB12},$$
$$\mu_{\mu\mu}^{exp}\leq6.8\times10^{-10}\mu_B, \qquad90\%\
C.L.\qquad [\mbox{LSND}]\ \cite{LBA01}.$$ For the $\tau$-neutrino,
the limits on $\mu_{\tau\tau}$ are less limitative (see, for
example \cite{tauB}), and the current upper bound on that is
$3.9\times10^{-7}\mu_B$.

The limits on the NCRs could be received from the studying the
elastic neutrino-electron scattering. For example, investigation
of this process at the TEXONO experiment results in the following
bounds on the NCR \cite{TSK15}
$$-2.1\times10^{-32}\ \mbox{cm}^2\leq(<r_{\nu_e}^2>)\leq
3.3\times10^{-32}\ \mbox{cm}^2.\eqno(4)$$

The AM of $1/2$-spin Dirac particle was introduced in the work
\cite{Y57} for a $T$-invariant interaction which violates
$P$-parity and $C$-parity, individually. Later in order to
describe this kind of interaction a more general characteristic,
the toroid dipole moment (TM) \cite{VM74}, was entered. It was
shown that the TM is a general case of the AM and at the
mass-shell of the viewed particle the both moments coincide. The
neutrino toroid interaction are manifested in scattering of the
neutrinos with charged particles. In so doing, the interaction
saves the neutrino helicity and gives an extra contribution, as a
part of the radiative corrections. In this regards, the AM is
similar to the NCR. Both quantities preserve the helicity in
coherent neutrino collisions, but have various nature. They define
the axial-vector (AM) and the vector (NCR) contact interactions
with an external electromagnetic field, respectively. From the
viewpoint of determining the NCR and AM the low-energy scattering
processes are of special interest (see, for example, Refs.
\cite{JL85,RC91}). The both neutrino interactions may have very
interesting consequences in various media. The possible role of
the AM in studying the neutrino oscillations was first specified
in Refs. \cite{VD98}. A point that should be also mentioned is
Ref. \cite{OMDR} in which the existence of the AM led to the
changing the flux of the solar electron neutrino during the
initial stage of the SF. Since phenomenology of the AM is
analogous to that of the NCR, the linkage between these quantities
must exist. In the SM for a zero-mass neutrino, the value of the
AM $a_{\nu}$ is connected with the NCR through the simple relation
(see, for example, \cite{AR00})
$$a^{\prime}_{\nu}=\frac{1}{6}<r_{\nu}^2>\eqno(5)$$
(the dimensionality of the AM in CGS system is
"$\mbox{length}^2\times\mbox{charge}$", that is to say,
$a_{\nu}=ea^{\prime}_{\nu}$ \cite{Y57}). However in the SM with
the massive neutrinos and in the case of the SM extensions this
relationship is violated \cite{MSD04}.

As for the CS magnetic fields, we shall assume that they are
nonhomogeneous, vortex and have the geometrical phase $\Phi(z)$
(twisting)
$$B_x\pm iB_y = B_{\bot}e^{\pm i\Phi(z)},\eqno(6)$$
where $\Phi(z)=\arctan(B_y/B_x)$. We notice that both  for the Sun
and for the Sun-like stars the reason of twisting is differential
rotation rates of their components and the global convection of
the plasma fluid. It should be recorded that configurations of the
solar magnetic field implying twisting nature are already being
discussed in the astrophysical literature for a long time (see,
for example \cite{NY71}). In Ref. \cite{VW90} the phase $\Phi$ was
introduced for the solar neutrino description for the first time.
Subsequently, in Ref. \cite{SM91} an account of this phase was
demonstrated. It should be remarked the works
\cite{AS91,ST91,ekh1993} which were devoted to the effects on
neutrino behavior in the twisting magnetic fields. For example, in
Ref. \cite{ekh1993} the neutrino beam traveling in the twisting
magnetic fields of the solar convective zone was considered and
some new effects (changing the energy level scheme, changing the
resonances location, appearing the new resonances, merging the
resonances and so on) were predicted. Assuming that the magnitude
of the twist frequency $\dot{\Phi}$ is determined by the curvature
radius $r_0$ of the magnetic field lines, $\dot{\Phi}\sim1/r_0$,
while $r_0$ has the order of $10\%$ of the solar radius, the
authors came to the following conclusion. To ensure that these new
effects will be observed the value of $\dot{\Phi}$ in the
convective zone should have the order of $10^{-15}$ eV.

Inasmuch as we shall take into investigation the interaction of
the neutrinos with the electromagnetic fields, the neutrino system
under study must contain both the left-handed and right-handed
neutrinos. By virtue of the fact that the right-handed Majorana
neutrinos are not sterile and interact as the right-handed Dirac
antineutrinos, we shall denote them as $\overline{\nu}_{lR}$. In
order to stress the sterility of right-handed Dirac neutrinos we
shall use for them the notation $\nu_{lR}$. So, in two-flavor
approximation the Majorana neutrino system will be described by
the function $(\psi^M)^T=(\nu_{eL},\nu_{\kappa
L},{\overline\nu}_{eR}, {\overline\nu}_{\kappa R})$ while for the
Dirac neutrinos we shall deal with the function
$(\psi^D)^T=(\nu_{eL},\nu_{\kappa L},\nu_{eR},\nu_{\kappa R})$. In
what follows to be specific, we shall reason $\kappa=\mu$.

To facilitate the evolution equation for the solar neutrinos we
transfer to the reference frame (RF) which rotates with the same
angular velocity as the transverse magnetic field. The matrix of
the transition to the new RF has the view
$$S=\left(\matrix{e^{i\Phi/2}&           0 & 0 & 0\cr
0 & e^{i\Phi/2}& 0 &0\cr 0 &0& e^{-i\Phi/2}& 0\cr 0&0&0&
e^{-i\Phi/2}}\right).\eqno(7)$$ In this RF the evolution equation
for the Dirac neutrinos is given by the expression
$$i\frac{d}{dz}\left(\matrix{\nu_{eL}\cr\nu_{\mu L}
\cr{\nu}_{eR} \cr{\nu}_{\mu
R}}\right)=\Big({\cal{H}}_0^D+{\cal{H}}^D_{int}\Big)
\left(\matrix{\nu_{eL}\cr\nu_{\mu L}\cr{\nu}_{eR}\cr{\nu}_{\mu
R}}\right),\eqno(8)$$ where
$$V_{eL}^{\prime}=V_{eL}+V^{\tilde{\delta}}_{ee},\qquad
V_{eL}=\sqrt{2}G_F(n_e-n_n/2),\qquad V_{\mu
L}=-\sqrt{2}G_Fn_n/2,$$
$$\Delta^{12}=\frac{\Delta m^2}{4E}=\frac{m^2_1-m^2_2} {4E},
\qquad
{\cal{A}}^{DL}_{ll^{\prime}}=\Big\{e\frac{<r_{\nu_{lL}\nu_{l^
{\prime}L}}^2>}{6}+a_{\nu_{lL}\nu_{l^{\prime}L}}\Big\}[\mbox{rot}\
{\bf{H}}(z)]_z,$$
$$\cos{2\theta}=c_{2\theta},\ \ \sin{2\theta}=s_{2\theta},
\qquad m_1=m_{\nu_e}\cos\theta-m_{\nu_{\mu}}\sin\theta,\qquad
m_{2}=-m_{\nu_e}\sin\theta+m_{\nu_{\mu}}\cos\theta,$$ $V_{eL}$
($V_{\mu L}$) is a matter potential caused by interaction of the
$\nu_{eL}$ ($\nu_{\mu L}$) neutrinos with the gauge bosons $W^-$
and $Z$, $V^{\tilde{\delta}}_{ee}$ is contribution to the matter
potential produced by the singly charged Higgs boson
$\tilde{\delta}^-$,
 $\theta$ is
a neutrino mixing angle in vacuum, $m_1$ and $m_2$ are mass
eigenstates, $\dot{\Phi}$ is the twisting frequency, $n_n$ ($n_e$)
is neutron (electron) density and the free Hamiltonian
$${\cal{H}}_0^D=\left(\matrix{-\Delta^{12}c_{2\theta}&\Delta^{12}s_{2\theta}&
0&0\cr\Delta^{12}s_{2\theta}&\Delta^{12}c_{2\theta}&0&0\cr
0&0&-\Delta^{12}c_{2\theta}&\Delta^{12}s_{2\theta}\cr
0&0&\Delta^{12}s_{2\theta}&\Delta^{12}c_{2\theta}\cr}
\right)\eqno(9)$$ describes oscillations in vacuum, while the
interaction Hamiltonian
$${\cal{H}}^D_{int}=\left(\matrix{V_{eL}+{\cal{A}}^{DL}_{ee}
-\dot{\Phi}/2 &{\cal{A}}^{DL}_{e\mu} & \mu_{ee}B_{\perp}
&\mu_{e\mu}B_{\perp}\cr{\cal{A}}^{DL}_{\mu e}&V_{\mu L}+
{\cal{A}}^{DL}_{\mu\mu}-\dot{\Phi}/2&\mu_{e\mu}B_{\perp}&
\mu_{\mu\mu}B_{\perp}\cr \mu_{ee}B_{\perp}
&\mu_{e\mu}B_{\perp}&\dot{\Phi}/2&0\cr
\mu_{e\mu}B_{\perp}&\mu_{\mu\mu}B_{\perp}&0&
\dot{\Phi}/2\cr}\right)\eqno(10)$$ covers interaction with medium.

When writing ${\cal{H}}_{int}^D$ we have taken into consideration
that the toroid interaction does not equal to zero in the external
inhomogeneous vortex magnetic field. In a concrete experimental
situation this field may be realized according to Maxwell's
equations as the displacement and conduction currents. The
universally adopted model of the SF is the magnetic reconnection
model \cite{KST11}. Owing to to it, a variable electric field
induced by variation of the CS magnetic field appears at the SF
initial phase. This field causes the conduction current which
takes on the appearance of a current layer aimed at the limiting
strength line which is common for the both CSs. So, in this case
the neutrinos are influenced by both the displacement current and
the conduction current.

For the Majorana neutrino case the evolution equation will look
like
$$i\frac{d}{dz}\left(\matrix{\nu_{eL}\cr\nu_{\mu L}\cr
\overline{\nu}_{eR} \cr\overline{\nu}_{\mu
R}}\right)=\Big({\cal{H}}_0^M+{\cal{H}}^M_{int}\Big)
\left(\matrix{\nu_{eL}\cr\nu_{\mu
L}\cr\overline{\nu}_{eR}\cr\overline{\nu}_{\mu
R}}\right),\eqno(11)$$ where $${\cal{H}}_0^M={\cal{H}}_0^D,$$
$${\cal{H}}_{int}^M=\left(\matrix{V_{eL}^{\prime}+{\cal{A}}^L_{ee}-\dot{\Phi}/2
&{\cal{A}}^L_{e\mu}&0&\mu_{e\mu}B_{\perp}\cr {\cal{A}}^L_{\mu
e}&V_{\mu L}+
{\cal{A}}^L_{\mu\mu}-\dot{\Phi}/2&-\mu_{e\mu}B_{\perp}&0\cr
0&-\mu_{e\mu}B_{\perp}&-V^{\prime}_{eL}+{\cal{A}}^R_{ee}+\dot{\Phi}/2&
{\cal{A}}^R_{e\mu}\cr \mu_{e\mu}B_{\perp}&0&{\cal{A}}^R_{\mu
e}&-V_{\mu
L}+{\cal{A}}^R_{\mu\mu}+\dot{\Phi}/2\cr}\right),\eqno(12)$$
$${\cal{A}}^L_{ll^{\prime}}=
\Big\{e[1-\delta_{ll^{\prime}}]\frac{<r_{\nu_{lL}\nu_{l^{\prime}L}}^2
>}{6}+
a_{\nu_{lL}\nu_{l^{\prime}L}}\Big\}[\mbox{rot}\ {\bf{H}}(z)]_z,$$
$${\cal{A}}^R_{ll^{\prime}}=\Big\{e[1-\delta_{ll^{\prime}}]
\frac{<r_{\overline{\nu}_{lR}\overline{\nu}_{l^{\prime}R}}^2>}
{6}-a_{\overline{\nu}_{lR}\overline{\nu}_{
l^{\prime}R}}\Big\}[\mbox{rot}\ {\bf{H}}(z)]_z.$$

Our next task is to investigate the resonance transitions of the
neutrino beam traveling the magnetic field of the CS which is the
source of the future SFs. Remember, that for the resonance
transition to occur, the following requirements must be met: (i)
the resonance condition must be carried out; (ii) the width of the
resonance transition must be nonzero; (iii) the neutrinos must
transits a distance comparable with the oscillation length.

In order to find the exact expressions for the resonance
conversion probabilities we should specify the coordinate
dependence of the quantities $n_e$, $n_n$ $B_{\perp}$,
$\dot{\Phi}$ and solve the evolution equation. Then, with the help
of the found functions $\nu_l(z)$, we could determine all
resonance conversion probabilities. Of course we shall be dealing
with numerical solution and, as a result, the physical meaning
will be far from transparent. Moreover, in the most general case
some of the resonance transitions may be forbidden. Therefore,
first we must establish which of these transitions are allowed and
which are forbidden. Further we shall follow generally accepted
scheme (see, for example, \cite{XS93,ekh1993}), namely, we shall
believe that all resonance regions are well separated what allows
us to put these resonances independent. As far as the twisting is
concerned, amongst existing the twisting models (see, for example
\cite{TKUB94}) we choose the simple model proposed in Ref.
\cite{CA92}
$$\Phi(z)=\frac{\alpha}{L_{mf}}z,\eqno(12)$$
where $\alpha$ is a constant and $L_{mf}$ is a distance on which
the magnetic field exists.

We begin with the resonant transitions of the $\nu_{eL}$ neutrinos
in the Dirac neutrino case. Here the $\nu_{eL}$ may experience
three resonance transitions. The first one is the
$\nu_{eL}\to\nu_{\mu L}$ (Micheev-Smirnov-Wolfenstein
--- MSW \cite{wol78,mikh85}) resonance transition. The
requirement of the resonance existence, the width of the
transition and the oscillation length are given by the expressions
$$\Sigma_{\nu_{eL}\nu_{\mu L}}=
-2\Delta^{12}c_{2\theta}+V^{\prime}_{eL}-V_{\mu L}+
{\cal{A}}^{DL}_{ee}-{\cal{A}}^{DL}_{\mu\mu}=0,\eqno(13)$$
$$\Gamma_{\nu_{eL}\nu_{\mu L}}\simeq
\frac{\sqrt{2}(\Delta^{12}s_{2\theta}+
{\cal{A}}^{DL}_{e\mu})}{G_F},\eqno(14)$$
$$L_{\nu_{eL}\nu_{\mu L}}=\frac{2\pi}{\sqrt{\Sigma_{\nu_{eL}
\nu_{\mu L}}^2+(\Delta^{12}s_{2\theta}+
{\cal{A}}^{DL}_{e\mu})^2}}.\eqno(15)$$ From Eqs.(14) and (15) it
follows that the oscillation length reaches its maximum value at
the resonance
$$\big(\Gamma_{\nu_{eL}\nu_{\mu L}})_{max}=\frac{2\sqrt{2}\pi}{G_F
[L_{\nu_{eL}\nu_{\mu L}}]_{max}},\eqno(16).$$ By virtue of the
fact that $(\Gamma_{\nu_{eL}\nu_{\mu L}})_{max}\simeq
3.5\times10^7$ cm, this resonance transition is realized before
the convective zone. Therefore it is unrelated to the SFs which
occurs in the solar atmosphere. That, in its turn, means that
under the MSW resonance the quantities ${\cal{A}}^{DL}_{ee}$,
${\cal{A}}^{DL}_{\mu\mu}$ and ${\cal{A}}^{DL}_{e\mu}$ play no
part. Estimation of the transition probability at the MSW
resonance could be fulfilled with the help of the Landau-Zener
formulae which in the case of the linear dependence of density on
distance is given by the expression \cite{LWJ}
$$P_{LZ}=\exp\Bigg\{\frac{-\pi\Delta m^2\sin^22\theta}{4E\cos2\theta
|d(\ln{n_e)}/dr|_{res}}\Bigg\}.\eqno(17)$$ Then using $P_{LZ}$ it
can be shown that the neutrino flux passing through the region of
this resonance must be reduced by about a factor of two as it was
verified by experiments \cite{Altman}.

Further we cross to treating the resonances of the $\nu_{eL}$
neutrinos traversing the CS magnetic field. In that case the
$\nu_{eL}$ neutrinos may experience the following resonance
transitions $$ \nu_{eL}\to{\nu}_{e R}, \qquad
\nu_{eL}\to{\nu}_{\mu R}.$$ The quantities characterizing the
$\nu_{eL}\to{\nu}_{e R}$ resonance transition are as follows
$$\Sigma_{\nu_{eL}{\nu}_{eR}}^D=V_{eL}+
{\cal{A}}^{DL}_{ee}-\dot{\Phi}=0.\eqno(18)$$
$$(L_{\nu_{eL}{\nu}_{eR}})_{max}
\simeq\frac{2\pi}{\mu_{ee}B_{\perp}}.\eqno(19)$$ The case, when
the term $A^{DL}_{ee}$ is negligible compared to $\dot{\Phi}$ and
the resonance condition amounts to
$$V_{eL}\simeq\dot{\Phi},\eqno(20)$$ is unreal.
Genuinely, in order for Eq.(18) to be satisfied, it is necessary
that the twisting magnetic field exists at a distance greater than
the solar radius. On the other hand, the currents producing the
inhomogeneous vortex magnetic field could reach the values of
$10^{-1}\ \mbox{A}/\mbox{cm}^2$. Then, for the CSs the quantity
$(a_{\nu_{eL}\nu_{eL}})[\mbox{rot}\ {\bf{H}}(z)]_z$ will have the
order of $10^{-30}$ eV and being negative it could compensate the
term of $V_{eL}$ in Eq.(18). In doing so the $\nu_{eL}\to{\nu}_{e
R}$ resonance may take place only in the corona.

We are coming now to the $\nu_{eL}\to{\nu}_{\mu R}$ resonance. The
pertinent expressions for this resonance will look like
$$-2\Delta^{12}c_{2\theta}+V_{eL}+
{\cal{A}}_{ee}^{DL}-\dot{\Phi}=0,\eqno(21)$$
$$(L_{\nu_{eL}{\nu}_{\mu R}})_{max}\simeq\frac{2\pi}{\mu_{e\mu}
B_{\perp}}.\eqno(22)$$ In the solar atmosphere, the term $V_{eL}$
in Eq. (21) are more less than $\Delta^{12}c_{2\theta}$ and play
no part. Analogously the quantity
$(a_{\nu_{eL}\nu_{eL}})[\mbox{rot}\ {\bf{H}}(z)]_z$ appears to be
also small compared with $\Delta^{12}c_{2\theta}$. Hence, the
$\nu_{eL}\to\nu_{\mu R}$ resonance may take place only at the cost
of the twisting, that is, when the relation
$$2\Delta^{12}
c_{2\theta}+\dot{\Phi}\simeq0.\eqno(23)$$ will be realized.

Let us determine the values of the parameter $\alpha$ which
provide the fulfilment of Eq.(23) for different solar neutrinos.
Assuming $\mu_{e\mu}=\mu_{ee}$, $B_{\perp}=10^5$ G and using for
$\mu_{ee}$ its upper bound $2.9\times10^{-11}\mu_B$ we obtain
$$-\alpha=\left\{\begin{array}{ll}
10^4,\qquad\mbox{for}\ E_{\nu}=0.1\ {\mbox{MeV}}\
(pp-{\mbox{neutrinos}}),\\[2mm]
10^2,\qquad\mbox{for}\ \ E_{\nu}=10\ {\mbox{MeV}}\
(^8B-{\mbox{neutrinos}}).
\end{array}\right.\eqno(24)$$
If the CS magnetic field increases to the value of $10^8$ G (as it
may be for the super flare case \cite{MLA}), the above mentioned
values of $|\alpha|$ are decreased by a factor of $10^3$. Thus it
becomes obvious that under the certain conditions the
$\nu_{eL}\to{\nu}_{\mu R}$ resonance transition may be observed.
The resonance condition (23) does not contain $n_e$ and $n_n$ and,
as a consequence, the $\nu_{eL}\to{\nu}_{\mu R}$ resonance may
occur both in the corona and in the chromosphere.

Now we introduce the quantity which characterizes weakening the
electron neutrino beam after traversing the resonant layer
$$\eta_{\nu_{eL}{\nu}_{\mu R}}=\frac{N_{i}-N_{f}}{N_{i}},$$ where
$N_{i}$ and $N_{f}$ are numbers of the $\nu_{eL}$ neutrinos before
and after the passage of the $\nu_{eL}\to{\nu}_{\mu R}$ resonance,
respectively. Again, to find the exact value of
$\eta_{\nu_{eL}{\nu}_{\mu R}}$ we should concretize the dependence
on distance of the quantities $n_e$, $n_n$, $B_{\perp}$ and solve
Eq.(8). However, to roughly estimate this quantity, it will
suffice to compare the resonance widths $\Gamma_{\nu_{eL}\nu_{\mu
L}}$ and $\Gamma_{\nu_{eL}{\nu}_{\mu R}}$, while taking into
account the value of $\eta_{\nu_{eL}\nu_{\mu L}}$. Calculations
result in
$$\eta_{\nu_{eL}{\nu}_{\mu R}}\simeq\left\{\begin{array}{ll}
2\times10^{-4},\qquad\mbox{when}\
\mu_{e\mu}=(\mu_{ee})_{upper}=2.9\times10^{-11}\mu_B,\
\ B_{\perp}=10^5\ G,\\[2mm]
0.12,\hspace{19mm}\mbox{when}\
\mu_{e\mu}=(\mu_{\mu\mu})_{upper}=6.8\times10^{-10}\mu_B,\
B_{\perp}=10^7\ G.
\end{array}\right.\eqno(25)$$
It should be noted that all the magnetic-induced resonances have
the resonance widths which are completely determined by the
quantity $\mu_{\nu_l\nu_{l^{\prime}}}B_{\perp}$. So, the foregoing
estimations remain valid for such resonance conversions. Then it
becomes evident that neutrino detectors utilizing the CE$\nu$NS
could display weakening the $\nu_{eL}$ beam already at
$B_{\perp}\geq10^7$ G.

Further we shall assume that the $\nu_{eL}$ beam has passed
through the MSW resonance before entering the CS magnetic field.
To put it another way, we shall deal with the beam which has been
weakened at the cost of the $\nu_{eL}\to\nu_{\mu L}$ resonance. In
the CS magnetic field the $\nu_{\mu L}$ neutrinos  produced due to
the $\nu_{eL}\to\nu_{\mu L}$ resonance may experience the
$\nu_{\mu L}\to{\nu}_{eR}$ and $\nu_{\mu L}\to{\nu}_{\mu R}$
resonance conversions. The corresponding resonance conditions will
look like
$$2\Delta^{12}c_{2\theta}+V_{\mu L}+
{\cal{A}}_{\mu\mu}^{DL} -\dot{\Phi}=0,\eqno(26)$$
$$V_{\mu L}+{\cal{A}}_{\mu\mu}^{DL}
-\dot{\Phi}=0.\eqno(27)$$ From Eq. (26) follows that the $\nu_{\mu
L}\to{\nu}_{\mu R}$ resonance appears to be allowed when
$$\dot{\Phi}<|V_{\mu L}|, |{\cal{A}}_{\mu\mu}^{DL}|, \qquad
 \mbox{and}\qquad
 V_{\mu L}+{\cal{A}}_{\mu\mu}^{DL}\simeq0.\eqno(28)$$
Note that in the conditions of the $\nu_{\mu L}\to{\nu}_{eR}$ and
$\nu_{eL}\to{\nu}_{\mu R}$ resonances a large value of
$\Delta^{12}c_{2\theta}$ could be compensated only by
$\dot{\Phi}$. However, the fulfillment of the condition (26)
requires that $\dot{\Phi}$ be positive, while the condition (21)
will be satisfied only if $\dot{\Phi}$ is negative.

So, the survival probabilities of the electron and muon neutrinos
are defined by the expressions
$${\cal{P}}_{\nu_{eL}\nu_{eL}}=1-({\cal{P}}_{\nu_{eL}{\nu}_{eR}}+
{\cal{P}}_{\nu_{eL}{\nu}_{\mu R}}), \qquad {\cal{P}}_{\nu_{\mu
L}\nu_{\mu L}}=1-({\cal{P}}_{\nu_{\mu L}{\nu}_{eR}}+
{\cal{P}}_{\nu_{\mu L}{\nu}_{\mu R}}),\eqno(29)$$ where the
contribution of the MSW-resonance has been eliminated for reasons
expounded above. From the foregoing expressions follows that the
AM and/or the NCR must be taken into account for the Dirac
neutrino case.

In what follows we shall discuss the oscillation picture for the
Majorana neutrinos. Here for $\nu_{eL}$ we shall deal with the
$\nu_{eL}\to\overline{\nu}_{\mu R}$ resonance transition only. The
relations being pertinent to this transition are as follows
$$-2\Delta^{12}c_{2\theta}
+V^{\prime}_{eL}+V_{\mu L}+{\cal{A}}_{ee}^L-{\cal{A}}_{\mu\mu}^R
-\dot{\Phi}=0\eqno(30)$$
$$\Gamma_{\nu_{eL}\overline{\nu}_{\mu R}}\simeq
\frac{\sqrt{2}(\mu_{e\mu}B_{\perp})}{G_F},\eqno(31)$$
$$L_{\nu_{eL}\overline{\nu}_{\mu R}}\simeq
\frac{2\pi}{\sqrt{\Sigma_{\nu_{eL}\overline{\nu}_{\mu R}}^2+
(\mu_{e\mu}B_{\perp})^2}}.\eqno(32)$$ In the solar atmosphere, the
terms $V^{\prime}_{eL}$, $V_{\mu L}$ and
$(a_{\nu_{eL}\nu_{eL}}+a_{\overline{\nu}_{\mu
R}\overline{\nu}_{\mu R}})[\mbox{rot}\ {\bf{H}}(z)]_z$ appear to
be small compared with $\Delta^{12}c_{2\theta}$. Therefore, the
$\nu_{eL}\to{\overline\nu}_{\mu R}$ resonance may take place only
at the cost of the twisting, that is, when the relation
$$2\Delta^{12}
c_{2\theta}+\dot{\Phi}=0.\eqno(33)$$ will be realized.

The oscillations picture will be incomplete, if we don't take into
consideration the oscillation transitions of the $\nu_{\mu L}$
neutrinos, which were produced due to the MSW resonance. In the CS
magnetic field these neutrinos may experience the $\nu_{\mu
L}\to\overline{\nu}_{eR}$ resonance conversion. The corresponding
resonance condition will look like
$$2\Delta^{12}c_{2\theta}-
\dot{\Phi}=0.\eqno(34)$$ From comparing of the obtained expression
with $\Sigma_{\nu_{eL}\overline{\nu}_{\mu R}}$ it follows that
when the $\nu_{eL}\to\overline{\nu}_{\mu R}$ resonance will be
forbidden then the $\nu_{\mu L}\to\overline{\nu}_{eR}$ resonance
is allowed, and vice versa. In that case the survival
probabilities for the $\nu_{eL}$ and the $\nu_{\mu L}$ beams will
be given by the expressions
$${\cal{P}}_{\nu_{eL}\nu_{eL}}=1-{\cal{P}}_{\nu_{eL}\overline{\nu}_{\mu R}}
,\qquad {\cal{P}}_{\nu_{\mu L}\nu_{\mu L}}=1-{\cal{P}}_{\nu_{\mu
L} \overline{\nu}_{e R}}.\eqno(35)$$ It should be also noted that
from the obtained equations it follows that the contributions
coming from the AM and the CNR can be safely neglected when the
neutrino is the Majorana particle.

\section{Conclusions}

The physics-based method of forecasting the high energy
Earth-directed SFs with the help of the neutrino detectors
utilizing the CE$\nu$NS has been investigated. Consideration has
been fulfilled within the left right symmetric model of the
electroweak interaction. It was assumed that the neutrinos possess
such multipole moments as the charge radius, the magnetic and
anapole moments. In so doing we have constrained by two flavor
approximation. The behavior of the neutrino beam traveling through
the CSs being the sources of the future SFs has been discussed. It
was speculated that the CS magnetic field is vortex,
nonhomogeneous and possesses twisting. For the geometrical phase
$\Phi(z)$ connected with magnetic field twisting $\dot{\Phi}(z)$
the simple model $\Phi=\exp[{\alpha z/L_{mf}]}$ has been employed.
Two neutrino beams are included into consideration, namely, the
$\nu_{eL}$ beam and the $\nu_{\mu L}$ beam which has been produced
after the passage of the MSW resonance. The investigations have
been completed both for the Majorana and for the Dirac neutrinos.
The evolution equation has been obtained in the Schrodinger-like
form and all the possible magnetic-induced neutrino resonance
transitions have been established.

It was shown that in the Majorana neutrino case  the decreasing of
the electron neutrino numbers is caused by the
$\nu_{eL}\to\overline{\nu}_{\mu R}$ resonance while decreasing the
muon neutrino numbers is connected with the $\nu_{\mu
L}\to\overline{\nu}_{eR}$ resonance. Both these resonances may be
in existence only when the CS magnetic field possesses twisting.
It should be stressed that under the fulfillment of the
$\nu_{eL}\to\overline{\nu}_{\mu R}$ resonance condition appearance
of the $\nu_{\mu L}\to\overline{\nu}_{eR}$ resonance is excluded,
and conversely. Also note, that for the Majorana neutrino the AM
and the NCF do not exert any influence on the values of the
oscillation parameters at the conditions of the Sun. In the
Majorana theory the right-handed neutrinos $\overline{\nu}_{\mu
R}$ and $\overline{\nu}_{eR}$ are physical particles whereas
detectors based on CE$\nu$NC are flavor-blind (at least with the
existing experimental technique). Then, since the total neutrino
flux is kept constant after traveling the resonances, the
detectors will not feel the change in the flavor composition of
the neutrino beam.

In the Dirac neutrino case the oscillation picture is richer. Here
we have the following resonances
$$\nu_{eL}\to\nu_{eR}, \qquad\nu_{eL}\to\nu_{\mu R}$$
for the electron neutrinos, and
$$\nu_{\mu L}\to\nu_{\mu R}, \qquad\nu_{\mu L}\to\nu_{eR}$$
for the muon neutrinos. It should be stressed that in this case
the $\nu_{eR}$ and $\nu_{\mu R}$ are sterile particles and they
cannot be recorded by detectors based on CE$\nu$NC. The
$\nu_{eL}\to\nu_{\mu R}$ and $\nu_{\mu L}\to\nu_{e R}$ -
resonances could be realized only in the magnetic field with
twisting, while the existing of the $\nu_{eL}\to\nu_{e R}$ and
$\nu_{\mu L}\to\nu_{\mu R}$ - resonances are only possible if
either or both the AM and the NCR have the values close to their
experimental bounds. For all the magnetic-induced resonances the
oscillation width depends on the quantity
$\mu_{ll^{\prime}}B_{\perp}$ which, in its turn, determines the
weakening of the neutrino beams traveling the CS magnetic field.
In this time, one should be expected that decreasing the electron
neutrino beam will be less than decreasing the muon neutrino beam,
since the upper bound on $\mu_{\mu\mu}$ is bigger than that on
$\mu_{ee}$. Then, for example, using the upper bound on
$\mu_{\mu\mu}$ and assuming $B_{\perp}=10^8$ G we shall get
weakening of the $\nu_{\mu L}$ beam being equal to 1.2. It is
obvious that the second generation detectors could observe such
weakening the neutrino flux. Note, that the sensitivity of the
measuring could be significantly improved when detectors with
different element compositions are exploited. Then the systematic
errors associated with the inaccuracy in determining the intensity
of the neutrino beam are mutually excluded.

So, the detectors based on CE$\nu$NC could be utilized for
forecasting the high-energy SFs only when the neutrino has a Dirac
nature. This also allows us to state that the observation of the
neutrino beams passing through the magnetic fields of the CSs,
which are the sources of the SF, will allow to determine the
neutrino nature.

The flares could occur in the Sun-like stars too. In that case the
high energy flares exhibit the serious danger to the crew of the
spacecraft. Consequently, the problem of forecasting the flare is
topical for the cosmic flights as well. Obviously that terrestrial
neutrino detectors will be of no avail when flying outside the
solar system. It is hoped that the problem could find its solution
with the help of neutrino detectors similar in design to the
RED-100 installed on a spacecraft.

\section*{Acknowledgments}
This work is partially supported by the grant of Belorussian
Ministry of Education No 20211660

\end{document}